\input{psfig.tex}
\documentstyle[12pt,aasms4]{article}
\lefthead{Peres et al.} 
\righthead{The Sun as an X-ray star II}
\begin{document}      
% Begin document proper

\title{The distribution of emission measure, and of heating budget, among the loops in the corona. }

\author{ G. Peres\altaffilmark{1,2}, S. Orlando\altaffilmark{3}, and F. Reale\altaffilmark{1}} 
\affil{$^{1}$ Dipartimento di Scienze Fisiche ed Astronomiche, Sezione di Astronomia, Piazza del Parlamento 1, 90134 Palermo, Italy; }
\affil{$^{3}$ Osservatorio Astronomico di Palermo, Piazza del Parlamento 1, 90134 Palermo, Italy} 

\author{R. Rosner} 
\affil{Depts. of Astronomy \& Astrophysics and Physics and Enrico Fermi Institute, The University of Chicago; Chicago,
Illinois, USA}

\altaffiltext{2}{e-mail: peres@oapa.astropa.unipa.it}

\begin{abstract}

The aim of this paper is to validate a methodology for connecting the
emission measure of individual solar coronal loops to the integrated
emission measure of the entire solar corona, and using this connection to
deduce the energetic properties of the corona, and then to show how this
methodology can be applied to observations of solar-like stellar coronae.
The solar validation is carried out by using spatially resolved X-ray
observations of the Sun obtained from the {\it Yohkoh} satellite. This
work is a further step in our effort to place the ``solar-stellar
connection" on a quantitative footing. In particular, we show how this
analysis procedure can be used in the context of archival {\it
Einstein}, {\it ROSAT} and EUVE
data, as well as Chandra and XMM Newton
data, as a complementary analysis tool to existing multi-thermal
component models.

\end{abstract}


\section{Introduction}

One of the core questions of stellar coronal physics is what determines
the intensity of both local heating of particular coronal structures
and the overall heating of the corona taken as a whole. In order to
answer this sort of question, one must look to both the Sun and other
(solar-like) stars: the Sun cannot tell us how the overall heating
rate of a stellar corona depends on the presumed ``control parameters"
(viz., stellar rotation) since most of them
are fixed, while in the absence of spatially resolved
observations, stars cannot inform us directly about local heating
processes. This point has motivated us in a series of recent papers to
examine how one might go about comparing solar and stellar observations
of what is presumably the same process -- the ``mechanical" heating of
stellar surface layers to produce an X-ray emitting atmosphere
(Peres, Orlando, Reale, Rosner, \& Hudson, 1997, 2000, Paper II; Orlando, Peres,
\& Reale 2000a; Paper I). The purpose of the present paper is to extend
this work by showing how our detailed understanding of the energetics of
individual solar coronal structures can be used to understand the
energy budget of solar-like stellar coronae.

The starting point of our analysis is the recognition that the X-ray
emitting solar corona appears to be entirely formed by plasma
magnetically confined in ``loops" (Vaiana et al.\ 1973), and that while
the emission from these loops can fluctuate substantially (e.g.,
Sheeley \& Golub 1979; Shimizu \& Tsuneta 1997), this emission is
sufficiently steady on the relevant sound crossing and radiative
cooling times that simple hydrostatic models suffice to describe them
most of the time.  Although the phenomenology of these coronal plasma
structures seems to be well understood, their heating mechanism remains
one of the riddles of coronal physics. Nevertheless, existing spatially
resolved observations are able to tell us about the overall energetics
of these structures, and to relate it to the ambient magnetic fields
(Vaiana \& Rosner, 1978; Title \& Schrijver, 1998). A central role in
our present discussion will be played by the piece-wise integrated
emission measure of individual loops,

\begin{equation}
\mbox{{\rm em}}(T) \equiv \int_{T}^{T + \Delta T} q(T') ~dT'~,
\end{equation}

\noindent
where $q(T)$ is the differential emission measure, defined as

\begin{equation}
q(T) \equiv n^2 \left( {{dT} \over {ds}} \right)^{-1}~,
\end{equation}

\noindent
$n$ is the plasma density, $s$ is a coordinate along the line-of-sight
to the coronal source, ($T_{\rm min}, T_{\rm max}$) is the temperature
interval within a given loop over which we integrate the coronal
emission, and $T_{\rm max}$ is the loop temperature maximum.  We shall
assume (with little loss of generality) that the minimum temperature
relevant to X-ray emitting loops ($T_{\rm min}$) is the same for all
quasi-steady loops. The emission measure has been long known to be a
highly useful observationally-constrained quantity for discussing the
energetics of stellar atmospheres\footnote{It is important to recognize
that while the emission measure is useful for discussions of the
overall energetics, its use is limited for discussions of detailed
heating mechanisms: the latter type of discussion requires knowledge of
the actual local plasma density, which is not available as long as the
spatial substructuring of solar coronal loops remains observationally
unconstrained.} (e.g., Athay 1966, Jordan 1980).  What makes the global
emission measure especially useful is the fact that the total solar
$EM$ often does not change significantly over time scales of 30 minutes
or more (Orlando et al.\ 2000b), with the obvious exception of flares
and similar rapidly evolving, large transients.  This observation
agrees with the notion that the global emission measure for the coronal
X-ray emitting matter, EM, is dominated by the steady X-ray emitting
loops.

Our plan is then as follows: following 
Paper I and Paper II, we will first discuss the
emission measure of individual solar coronal loops in the specific
context of {\it Yohkoh} X-ray images (\S II), and show how these
individual emission measures can be simply summed by taking advantage of
elementary analytical properties of loop emission measures to obtain the
overall coronal emission measure. Finally, demonstrate how properties of
this integrated emission measure can be used to make deductions about the
heating of the loop structures themselves (\S III). This lattermost step
then allows us to apply our analysis to the spatially unresolved
observations of solar-like stellar coronae (\S IV).

\section{Connecting the global coronal emission measure with the
emission measure of single loop structures}

The standard Yohkoh data analysis yields pixel maps of temperature
and of emission measure from the ratio of two images taken with different
filters.  From those maps one derives the global distribution of emission
measure versus temperature, EM$(T)$: one divides the temperature range
into bins of width $\Delta T$, and sums the emission measure of all the
pixels belonging to the same temperature bin (Papers I, II). The EM$(T)$
distribution is then the emission measure of the plasma at temperature $T$
summed over the temperature range $\Delta T$, or (similar to
Eq. 1)
\begin{equation}
\mbox{{\rm EM}}(T) = \int_{T}^{T+\Delta T} Q(T') ~dT'~,
\end{equation}
\noindent
where $Q(T)$ is defined as in Eq. (2); we shall assume that $T$ varies
in the range ($T_{\rm min}, T_{\rm Max}$), where $T_{\rm min}$ is the minimum
temperature of material within loops contributing to X-ray emission, and
$T_{\rm Max}$ is the maximum value of all values of the loop maximum
temperature $T_{\rm max}$,
\begin{equation}
T_{\rm Max} \equiv {\rm max}\{T_{\rm max}\}~.
\end{equation}
It is worth noting that $Q(T)$ is a presumably smooth, continuous,
function while EM$(T)$ is a discretized function in the form of a
histogram.
In our analysis we take $\Delta \log T$ constant, with 29 bins in the
range $5.5 \le \log T(\mbox{K}) \le 8$, the nominal range of
Yohkoh/SXT, i.e., $\Delta \log T = 2.5/29 \sim 0.09$. Since $\Delta
\log T$ is small, we approximate $\Delta T \approx T\Delta \log T \ln
10 \approx \eta T$, with $\eta \approx 0.2$. In this section, we shall
connect the emission measure distribution EM$(T)$ to the distribution
of emission measures for individual loops.

\subsection{The emission measure distribution for an individual loop}

First, let us turn to the emission measure of individual loops. We
recall from earlier work (viz., Rosner, Tucker and Vaiana, 1978, in
the following RTV; Maggio and Peres 1996) that the functional
dependence of the emission measure of individual loops, {\it em}, on
temperature depends only on the maximum temperature of the given loops,
apart from a (normalizing) multiplicative factor; furthermore, we
recall that the emission measure of an individual steady loop, em$(T)$,
and the ascending part of the global emission measure, EM$(T)$, of the
entire solar corona both obey a simple power law temperature
dependence, starting from the minimum temperature ($T_{\rm min}$), of
the form

\begin{equation}
T^{3/2}~ .
\end{equation}

In the case of the global
emission measure, this was already recognized by Athay (1966; also
Jordan 1980, Ventura et al.\ 1998), while Jordan\ (1980) --
studying the very same emission measure -- conjectured that EM$(T)$ may
be interpreted as due to the sum of contributions from many individual
loops. More recently Laming, Drake and Widing (1995) have derived
the whole solar disk using the EUV data collected by Malinovsky and
Heroux (1973); the emission measure they derive is fully consistent with
the above functional form.

The analysis to follow below shows not only that Jordan (1980)
was correct, but that this synthesis procedure can be
inverted in order to derive the properties of the underlying
contributing loops from the integral (global) emission measure.
Finally, we note that while the emission measure em$(T)$ of an
individual loop terminates at the maximum loop temperature $T_{\rm
max}$, the global EM$(T)$ continues for temperatures beyond that at
which its maximum value is attained, with a functional form that also
can be approximated by a power law of the form \begin{equation}
T^{-n}~, \end{equation} where $n$ is typically much larger than $3/2$.

Let us now consider the synthesis of individual loop emission measures.
For convenience, we write the differential emission measure for a given
loop as
\begin{equation}
q(T) = bT^{\gamma} ~;
\end{equation}
numerical solutions of the one-dimensional hydrostatic loop equations
show that this power law scaling (with $\gamma \approx 1/2$) holds well
except for loops whose length far exceeds their scale height, or for
loops with highly non-uniform local heat deposition. Since we shall not
concern ourselves with these more complex circumstances (see below), we
can use this power law relation to integrate as follows:

\begin{equation}
\mbox{{\rm em}}(T) = \int_{T}^{T+\Delta T} bT^{\prime \gamma}
~dT' =
\frac{b}{\gamma+1}T^{\gamma+1}\left((1+\eta)^{\gamma+1}-1\right) \sim
b \eta T^{\gamma+1} ~.
\end{equation}

\noindent
Since $T_{\rm max}$ is the maximum loop temperature, we have from the
above equation, em$(T_{\rm max}) = b \eta T_{\rm max}^{\gamma+1} =
bT_{\rm max}^{\gamma} \eta T_{\rm max} = q(T_{\rm max}) \eta T_{\rm
max}$. In our specific treatment, em$(T)\propto T^{\beta} \sim
T^{3/2}$; thus we find $\gamma = 1/2$, so that the differential
emission measure $q(T)$ for a given loop can be expressed in terms of
the em$(T_{\rm max})$,

\begin{equation}
q(T) = bT^{\gamma} = \frac{\mbox{\rm em}(T_{\rm max})}{\eta T_{\rm max}}
\left(\frac{T}{T_{\rm max}}\right)^{\beta - 1}=
\frac{\mbox{\rm em}(T_{\rm max})}{\eta T_{\rm max}}
\left(\frac{T}{T_{\rm max}}\right)^{1/2} 
\label{eq3}
\end{equation}

\subsection{The emission measure distribution of loops with the same
maximum temperature}

As pointed out by many earlier authors, most of the X-ray emission from
the solar corona comes from plasma that lies within a (coronal)
pressure scale height of the surface; and since most coronal loops appear
to have lengths that are equal to, or less than, the pressure scale height
corresponding to their maximum temperature, this also means that the bulk
of coronal X-ray emission derives from loops with size less than a
pressure scale height. (This is not to say that loops that violate
these conditions do not exist: such loops do exist, but we argue that
they do not make a significant contribution to the integrated coronal
X-ray flux.)

As a consequence, we can take advantage of an extremely useful fact
about loops with roughly constant pressure: the one-dimensional
equations governing the physics of hydrostatic loops of this type are
invariant under changes of the loop field-line spatial scale (RTV;
Maggio \& Peres 1996), so that -- aside from a possible multiplicative
constant factor -- the functional dependence of the loop emission
measure distribution versus temperature, $\mbox{\rm em}(T)$, does not
depend on the loop length, but only on the loop maximum temperature
$T_{\rm max}$. Thus, adding the loop emission measures for two or more
loops with the same maximum temperature yields exactly the same
functional form versus temperature, apart from the normalization.

With this enormous simplification, we can group all loops of the same
maximum temperature $T_{\rm max}$ together, irrespective of their length
(and plasma pressure); 
this group may be called a $T_{\rm max}$-equivalence class of
loops (obviously indexed by the maximum loop temperature for that
equivalence class). Note that, since the product {\em plasma pressure
$\times$ loop length} determines the coronal temperature (RTV), loops
with very different pressure and length may belong to the same
$T_{\rm max}$ equivalence class.\ \ Thus, we can use the results of the
previous subsection, Eqs. (8) and (9), to write the
emission measure for the entire equivalence class of loops with a given
maximum temperature $T_{\rm max}$, Em($T_{\rm max}$), in the form
\begin{equation}
{\rm Em}(T,~ T_{\rm max}) = q_M ~(T/T_{\rm max})^{\beta} ~,
\end{equation}
where $q_M$ is the normalizing factor for the emission measure (the
maximum value, occurring at $T = T_{\rm max}$), and $\beta$ is always
approximately $3/2$.
We have also written ${\rm Em}(T,~ T_{\rm max})$ to state
explicitly its dependence on both $T$ and $T_{max}$.

\subsection{Summing the emission measure of all coronal loops}

We now model the emission measure of the entire corona as the sum
of a multitude of static loops of different maximum temperature. If we
define $f(T_{\rm max})~dT_{\rm max}$ as the probability that a loop has a
maximum temperature $T_{\rm max}$, such that $\int f(T_{\rm max})~dT_{\rm
max} = 1$, and $N$ is the total number of loops\footnote{Equivalently one
may formulate the entire treatment in terms of the amount of solar surface
covered by the loops with temperature maximum $T_{\rm max}$ and emission
measure of the loop per unit area, but this would not allow us to
disentangle the dependence on pressure (or length) from the area.}, then
the emission measure vs. $T$ of all the loops of equal maximum temperature $T_{\rm
max}$, Em($T, T_{\rm max}$), is given by
\begin{equation}
{\rm Em}(T, T_{\rm max}) = N f(T_{\rm max})~em(T_{\rm max}) ~(T/T_{\rm
max})^{3/2},
\end{equation}
i.e., it too is proportional to $T^{3/2}$ up to the temperature maximum;
the corresponding maximum emission measure is thus
$~N~f(T_{\rm max})~em(T_{\rm max})$.

The overall emission measure of the corona is due to the sum of all
equivalence classes of $T_{\rm max}$ loops present in the
corona up to the largest coronal temperature, here for simplicity
indicated
as $\infty$; this is simply

\begin{equation}
{\rm EM}(T) =~~\int_T^{\infty} ~~ dT_{\rm max}~N~f(T_{\rm max})~em(T_{\rm
max}) ~(T/T_{\rm max})^\beta~.
\label{EMsumloop}
\end{equation}

Thus, the global emission measure at temperature $T$ is the sum of the
emission measure of all the loops whose maximum temperature $T_{\rm
max} \ge T$. Fig. \ref{EMsum} shows an idealized example of such a sum
for a distribution of loops with different $T_{\rm max}$: we hypothesize
a simplified coronal loop population made of six different equivalence
classes and derive their EM(T) from Eq. \ref{EMsumloop}. Since the
contribution of each $T_{\rm max}$ loop equivalence class is $\propto
T^\beta$, the total emission measure distribution is also $\propto
T^\beta$, up to the lowest value of $T_{\rm max}$ of all the loop
equivalence classes. This will be proved more precisely in the
following. The inset of Fig. \ref{EMsum} shows the $~N~f(T_{\rm
max})~em(T_{\rm max})$ of the six loop equivalence classes used for this
example.  More generally, the shape of the global emission measure
distribution is determined by the relative contribution of each
equivalence class of loops to the overall emission measure (Drake et
al.\ 2000).  For example, while the EM$(T)$ distribution of the Sun
has, typically, only one maximum, the EM$(T)$ distribution of very
active stars may show two maxima (e.g., Griffiths \& Jordan 1998).  We
will comment on this point in the last section.

\subsection{The coronal EM$(T)$ derived from observations of the solar
corona}

Fig. \ref{EM1} shows the EM$(T)$ derived from a coronal observation
made with Yohkoh/SXT. The ascending part of the curve, up to its peak,
at the temperature
$T_P\approx 2 \times 10^6$ K, is reasonably well approximated by a
$T^{3/2}$ power law (cf.\ also Papers I, II, and Orlando et al.\ 2000b)
up to the peak value, $EM_P$ ($\approx 3.5 \times 10^{49}$
cm$^{-3}$). This power law behavior for the ascending part of the
emission measure is present in most phases of the solar cycle; Paper II
and Orlando et al.\ (2000b) show several examples of this behavior. If
one examines the entire set of Yohkoh/SXT observations along the solar
cycle, one finds that $T_P$ varies only by a small factor, while the
amplitude of the emission measure at $T=T_P$, $EM_P = EM(T_P)$, varies
by several orders of magnitude; both these variations
are larger at the time of solar maximum.

For $T>T_P$, the emission measure declines, and again closely follows a
single power law of the form $T^{-\alpha}$, with $\alpha \sim 3.5$ (Fig.\ 
\ref{EM1}). This behavior holds for all cases of the quasi-steady solar
corona that we have studied, although the power law index $\alpha$ varies
as the cycle proceeds: $\alpha$ is larger (steeper power law) near solar
minimum (cf. Paper II). The close resemblance between the synthetic
emission measure distribution shown in Fig.\ \ref{EMsum} and the
observationally-derived emission measure distribution shown in Fig.\ 
\ref{EM1} suggests strongly that the observed distribution can indeed be
regarded as a result of the superposition of quasi-static loops
characterized by a distribution of maximum loop temperatures; in the
following, we will then simply presume that this is indeed the case.

\subsection{Deriving the contribution of the various loop equivalence
classes}

We now show how the contribution of each equivalenced class of
loops, e.g., the unknown expression $N~f(T_{\rm max})~em(T_{\rm max})$
appearing in Eq. (12), can be recovered from an analysis of
the observed EM$(T)$: in Appendix A we show that this expression can be
obtained by using Eq. \ref{EMsumloop} to give
\begin{equation}
N~f(T_{\rm max})~em(T_{\rm max})=~ \beta \frac{\mbox{EM}(T_{\rm
max})}{T_{\rm max}} -
\frac{d~\mbox{EM}(T_{\rm max})}{dT_{\rm max}} ~.
\label{Nfqeqn}
\end{equation}
Now, since 
\begin{equation}
\mbox{EM}(T) \approx \left\{
\begin{array}{ll}
\displaystyle
\mbox{EM}_M \left(\frac{T}{T_P}\right)^\beta & ~~~\mbox{for  } T\leq T_P \\
\\
\displaystyle
\mbox{EM}_M \left(\frac{T}{T_P}\right)^{-\alpha} & ~~~\mbox{for  }
T>T_P
\end{array} \right.
\end{equation}
we obtain immediately
\begin{equation}
Nf(T_{\rm max})~em(T_{\rm max}) \approx \left\{
\begin{array}{ll}
\displaystyle
0 & \mbox{for  } T_{\rm max}<T_P\\
\\
\displaystyle
(\alpha+\beta) \frac{\mbox{EM}_M}{T_P}
\left(\frac{T_{\rm max}}{T_P} \right)^{-(\alpha+1)}  & \mbox{for  }T_{\rm
max}>T_P
\end{array} \right.
\label{loop_contrib}
\end{equation}
\noindent
The dashed line in Fig. \ref{Nfq} shows the function $Nf(T_{\rm max})
em(T_{\rm max})$ derived from the emission measure distribution EM$(T)$
shown in Fig. \ref{EM1}, i.e. computed from Eq. \ref{loop_contrib}
with $\beta= 1.5$ and $\alpha= 3.5$. Eq. \ref{loop_contrib} and Fig.
\ref{Nfq} show that $N~f(T_{\rm max})~em(T_{\rm max})$ is a sharply
peaked function around $T_P$, that there are no (or, more
realistically, just very few) loops with $T_{\rm max} < T_P$, and that
the distribution itself decreases for $T_{\rm max} > ~ T_P$ as a power
law whose index is smaller by one (i.e., is steeper) than that of
EM$(T)$.

Fig. \ref{Nfq} also shows the results of an analysis of the EM($T$),
using Eq. \ref{Nfqeqn} but taking into account the departure from power
laws of the ascending and of the descending parts of EM(T), and
considering the effect of error bars and the inherently non-linear
characteristics of the analysis.  We have made a Monte-Carlo
calculation as follows: first for each data point of EM($T$) we have
generated a random value from a Gaussian distribution centered on the
value itself with $\sigma$ given by the error bar; then applying Eq.
\ref{Nfqeqn} to this new EM(T) distribution we obtain a $N~f(T_{\rm
max})~em(T_{\rm max})$. We have thus repeated the Monte-Carlo
calculation 1000 times and determined, for each $T_{max}$, the median
value (marked with a diamond) and the bounds enclosing  68\% and 90\%
of the distribution of $N~f(T_{\rm max})~em(T_{\rm max})$ values
obtained with the simulation. These two bounds are shown as error bars
of different lengths on each point in Fig. \ref{Nfq}. The $N~f(T_{\rm
max})~em(T_{\rm max})$ distribution obtained in this way is remarkably
close to the results of Eq. \ref{loop_contrib}, albeit the latter
somehow underestimates the contribution of loops in the descending
part, especially hot ($T_{max} \approx 10^7$ K) loops, and of loops
with $T_{max} \approx 10^6$ K.  The results well below $10^6$ K are
consistent with no significant contribution from soft X-ray emitting
loops, however with large error bars.

For the analytical derivation to follow we will use the piecewise
power law description made above, for ease of calculation and because it
can help us to gain insight into the results.  Also, we will discuss
again the limits of our approximations for $T_{\rm max} < T_P$ in the
last section.

As an aside, we note (from Eq. (11)) that each equivalence class of
loops with the same maximum temperature $T_{\rm max}$ will contribute
to the {\em total} emission measure as follows:

\begin{eqnarray}
\lefteqn{\mbox{Em}_{tot}(T_{\rm max})= \int_{T_{\rm min}}^{T_{\rm max}}
dT~N~f(T_{\rm
max})~em(T_{\rm max}) ~\left(\frac{T}{T_{\rm max}}\right)^{\beta} }
\nonumber
\\
 & ~~~~~~~~~~ \displaystyle
\approx \frac{\alpha + \beta}{\beta + 1}~\mbox{EM}_M ~
\left(\frac{T_{\rm max}}{T_P}\right)^{-\alpha} 
\end{eqnarray}

\noindent
where $T_{\rm min} \sim 2 \times 10^5$ K is the lowest coronal temperature we
consider, and since $T_{\rm min} \ll T_{\rm max}$ the term containing $T_{\rm min}$ is
negligible.

\section{Constraints on coronal heating}

The fact that EM$(T)$ can be expressed as a sum over the emission
measure distribution of many independent static (or quasi-static) loops
has important implications on the heating of these structures.

\subsection{The heating budget of loops with the same maximum temperature}

If $\mbox{\~Q}(T)$ is the differential emission measure per unit temperature interval
at temperature $T$ of a whole class of loops of the same maximum temperature,
$T_{\rm max}$, and $P(T)$ is the radiative loss per unit emission
measure, then the total radiative losses are given by
\begin{equation}
R=~ \int_{T_{\rm min}}^{T_{\rm max}} {dT ~\mbox{\~Q}(T)~P(T)} ~.
\end{equation}
Since 
\begin{equation}
\mbox{\~Q}(T)= ~N~f(T_{\rm max})~em(T_{\rm max})\times \frac{1}{\eta
T_{\rm max}}\left(\frac{T}{T_{\rm max}}\right)^{\beta - 1} ~,
\end{equation}
\noindent
in analogy to what was discussed in \S 2, and $P(T) \approx 
P_0 T^{-1}$ in the range $2 \times 10^5 ~\mbox{K}<~T~<10^7 ~\mbox{K}$,
with $P_0 \sim 2.02 \times 10^{-16}$ erg s$^{-1}$ cm$^{3}$ K (from the
MEKAL spectral model; Mewe, Lemen, \& van den Oord 1986; Kaastra
1992; Mewe, Kaastra \& Liedahl 1995 and references therein),
\begin{eqnarray}
\lefteqn{R= \frac{N~f(T_{\rm max})~em(T_{\rm max})}{\eta}~\frac{P_0}{T_{\rm
max}^{\beta}} \int_{T_{\rm min}}^{T_{\rm max}} T^{\beta -2}~dT }
\nonumber
\\
 &  \displaystyle \approx \frac{N~f(T_{\rm max})~em(T_{\rm max})}{\eta
(\beta - 1)} ~P_0 ~T_{\rm max}^{-1}~,~~~~~~~~~~~~~~~~
\end{eqnarray}
\noindent
where we have neglected $T_{\rm min}^{\beta -1}$ since $T_{\rm min} \ll T_{\rm max}$.

These radiative losses are approximately one half of the total energy
input to the loop, the other half being conducted towards the lower
transition region, where it is radiated away (cf., Vesecky, Antiochos, \&
Underwood 1979). The total amount of heat delivered within loops
with temperature maximum $T_{\rm max}$ is thus given by
\begin{equation}
h(T_{\rm max}) \approx  \frac{2}{\eta(\beta - 1)}~N~f(T_{\rm
max})~em(T_{\rm max}) ~P_0~T_{\rm max}^{-1} 
\label{eq12}
\end{equation}
\noindent
Therefore, using Eq. \ref{loop_contrib}
\begin{equation}
h(T_{\rm max})\approx \left\{
\begin{array}{ll}
0 & \mbox{for } T_{\rm max}<T_P \\
\\
\displaystyle \frac{2(\alpha+\beta)}{\eta (\beta - 1)}
\frac{\mbox{EM}_M~P_0}{T_P^2}~
\left(\frac{T_{\rm max}}{T_P} \right)^{-(\alpha+2)} & \mbox{ for } T_{\rm
max}\ge T_P~.
\end{array}\right.
\label{heat_budget}
\end{equation}

Fig. \ref{h(T)} shows the heating budget $h(T_{\rm max})$ for each
equivalence class of loops characterized by the temperature maximum
$T_{\rm max}$, as derived from the function $N f(T_{\rm max}) em(T_{\rm
max})$ of Fig. \ref{Nfq}; recall that $\beta
\approx 3/2$ and that, for the case shown in Fig. \ref{EM1}, $T_P \approx
1.7 \times 10^6$K, $EM_P \approx 3.5 \times 10^{49} $ cm$^{-3}$
and
$\alpha \approx 3.5$. Thus, the total coronal loop heating rate up to a
reference temperature $t$ is given simply by
\begin{equation}
H(t) \approx 2 R = 2~\int_{T_{\rm min}}^t dT ~Q(T) ~P(T) ~.
\label{heat_cpt}
\end{equation}
\noindent
Our aim is now to evaluate this equation for the reference temperature $t$,
and then to allow $t$ to increase indefinitely in order to obtain the total
coronal heating rate. On the basis of Eq.
\ref{eq3}, $Q(T) = Q_M (T/T_P)^{\xi-1}$, $Q_M = \mbox{EM}_M/\eta T_P$ and,
for generality, we indicate with $\xi$ the index of the power laws in
the ascending and in the descending part of $EM(T)$:
\begin{equation}
\xi = \left\{
\begin{array}{ll}
\beta        &  \mbox{   for  } T<T_P \\
-\alpha  &  \mbox{   for  } T>T_P
\end{array}
\right .
\end{equation}
\noindent
From Eq. \ref{heat_cpt},
\begin{equation}
H(t) = 2~\int_{T_min}^t dT ~Q(T) ~P(T) =
2\frac{\mbox{EM}_M}{\eta T_P} \int_{T_min}^{t} 
\left (\frac{T}{T_P} \right)^{\xi-1} P(T) ~dT = 
2\frac{\mbox{EM}_M}{\eta T_P} P_0 \int_{\tau_min}^{\tau'}
\tau^{\xi-2} d\tau
\end{equation}
\noindent
where $\tau = T/T_P$, $\tau_{\rm min} = T_{\rm min}/T_P$, $\tau' = t/T_P$,
and
\begin{equation}
H(t) = \frac{2\mbox{EM}_M P_0} {\eta T_P}\times
\left\{
\begin{array}{ll}
\displaystyle
\frac {(t/T_P)^{\beta -1} - (T_{\rm min}/T_P)^{\beta -1}}{\beta -1} &
\mbox{for~~~~} t\le T_P \\
\\
\displaystyle
\frac {1 - (T_{\rm min}/T_P)^{\beta -1}}{\beta -1}
+ \frac{1-(t/T_P)^{-(\alpha+1)}}{\alpha+1} &
\mbox{for~~~~} t> T_P
\end{array}\right.
\end{equation}
\noindent
which for $t \gg T_P$ yields the total heating budget
\begin{equation}
H_{tot} = \frac{2\mbox{EM}_M P_0}{\eta T_P}
\left[\frac {1 - (T_{\rm min}/T_P)^{\beta -1}}{\beta -1}
+\frac{1}{\alpha+1}\right]
\end{equation}
\noindent
and for $T_{\rm min} \ll T_P$
\begin{equation}
H_{tot} \approx \frac{2\mbox{EM}_M P_0}{\eta T_P}
~\left( \frac{1}{\beta -1} + \frac{1}{\alpha +1} \right)~.
\end{equation}
\noindent
Fig. \ref{H(t)} shows the curve $H(t)$ for the emission measure
distribution $EM(T)$ of Fig. \ref{EM1}. The above equation shows that the
entire coronal heating budget is approximately proportional to the peak
value
of the emission measure distribution ($EM_P$), and inversely proportional to
the temperature of maximum emission measure ($T_P$), with a rather mild
dependence on the indices of the ascending and descending power laws,
$\beta$ and $ - \alpha$, respectively, which just enter via the factor
\begin{equation} 
\frac{2}{k} \left( \frac{1}{\beta -1} + \frac{1}{\alpha +1} \right) ~;
\end{equation}
\noindent
the value of this factor is approximately 20 for the specific case
shown in Fig. \ref{EM1}.

For the specific case shown in Fig. \ref{EM1}, $H_{tot}\approx 9.6 \times
10^{28}$ ergs s$^{-1}$, i.e., approximately twice the total radiative
losses we found in paper II for the same emission measure distribution;
this is of course not a surprise since we built in this fact in our
assumption regarding the evaluation of $H$ immediately preceeding Eq.
\ref{eq12}.

\subsection{Average heating rate per unit volume within loops}

The total heat deposited in a set of loops characterized by maximum
temperature $T_{\rm max}$ can also be expressed as 
\begin{equation}
h(T_{\rm max})=  \sum_j E_{Hj}~L_j~a_j ~,
\label{h_setloop}
\end{equation}
\noindent
where $E_{Hj}$ is the average heating rate per unit volume, assumed
uniform inside the loop for simplicity, $a_j$ is the area covered by the two footpoints on the solar (or
stellar) surface and $L_j$ is the semilength of the loops with $T_{\rm
max}$ as temperature maximum. The index $j$ runs over all loops with
maximum temperature $T_{\rm max}$.

We now consider the two scaling laws for the plasma confined inside a
hydrostatic loop (RTV):
\begin{equation}
T_{\rm max} = 1.4 \times 10^3 (p L)^{1/3} 
\label{tmax}
\end{equation} 
\begin{equation} 
E_H= 10^5 ~p^{7/6} ~ L^{-5/6} 
\label{heating}
\end{equation}
\noindent
where $p$ is the coronal pressure (by assumption, essentially constant
within loops shorter than the pressure scale height), and $E_H$ is the
average coronal heating per unit volume inside such loops. These equations
allow us to solve for the pressure in terms of the loop length and maximum
temperature, 
\begin{equation}
p = \frac{1}{L}\left(\frac{T_{\rm max}}{1.4 \times 10^3}\right)^3 
\label{eq24}
\end{equation}
\noindent
and, upon substituting Eq. \ref{eq24} into Eq. \ref{heating}, one
obtains
\begin{equation}
E_H= \left\{
\begin{array}{ll}
0 & \mbox{for } T_{\rm max} < T_P\\
\\
\displaystyle 10^5 \left(\frac{T_{\rm max}}{1.4 \times
10^3}\right)^{7/2}~L^{-2} &
\mbox{   for } T_{\rm max}\ge T_P~.
\end{array}\right.
\label{eq25}
\end{equation}
\noindent
where we have taken into account the fact that, in practice, very few loops
exist with $T_{\rm max} < T_P$. We can now solve for $L$ for $T_{\rm
max}\ge T_P$ from the above equation, and substitute into Eq.
\ref{h_setloop} to obtain, for $T_{\rm max}\ge T_P$
\begin{equation}
h(T_{\rm max})= 9.9\times 10^{-4}~T_{\rm max}^{7/4} \sum_j E_{Hj}^{1/2}~a_j
\approx 10^{-3} T_{\rm max}^{7/4} \sum_j E_{Hj}^{1/2}~a_j ~.
\end{equation}
\noindent
Since this expression does not depend on $L$, it therefore applies to all
loops with the same maximum temperature $T_{\rm max}$. Equating the
right-hand sides of this equation and Eq. \ref{heat_budget}, for
$T_{\rm max}\ge T_P$
\begin{equation}
\frac{2(\alpha + \beta)}{\eta (\beta - 1)}\frac{\mbox{EM}_M~P_0}{T_P^{2}}
\left(\frac{T_{\rm max}}{T_P}\right)^{-(\alpha+2)}=
10^{-3} T_{\rm max}^{7/4} \sum_j E_{Hj}^{1/2}~a_j
\end{equation}
therefore
\begin{equation}
\sum_j E_{Hj}^{1/2}~a_j = \left\{
\begin{array}{ll}
0 & \mbox{for } T_{\rm max} < T_P\\
\\
\displaystyle 2\times 10^3  \left[\frac{EM_P 
P_0}{\eta} \frac{\alpha + \beta}{\beta -1} {T_P}^\alpha
\right] T_{max}^{-(\alpha + 3.75)} &
\mbox{   for } T_{\rm max}\ge T_P~.
\end{array}\right.
\end{equation}
\noindent
For the EM$(T_{\rm max})$ shown in Fig. \ref{EM1}, $\alpha \approx 3.5$;
using the typical value $\beta= ~1.5$, we find that
\begin{equation}
\sum_j E_{Hj}^{1/2}~a_j = \left\{
\begin{array}{ll}
0 & \mbox{for } T_{\rm max} < T_P\\
\\
\displaystyle 2\times 10^4  \left[\frac{EM_P 
P_0}{\eta} {T_P}^{3.5} \right] T_{\rm max}^{-7.25} &
\mbox{   for } T_{\rm max}\ge T_P~.
\end{array}\right.
\label{eq29}
\end{equation}
Fig. \ref{Ea} shows, as an example, the expression $\sum_j
E_{Hj}^{1/2}~a_j$ corresponding to the emission measure distributions
EM$(T_{\rm max})$ of Fig. \ref{EM1}.

Now, in order to derive the volumetric heating rate $E_H$ one needs to
know the coefficients $\{a_j\}$ which, however, cannot be derived from the
emission measure itself, nor can they be derived from stellar
observations.

Finally, we note that the power law dependence on $T_{\rm max}$ in Eq.
\ref{eq29} is rather steep. However,  common experience with coronal loops
shows that hotter loops (higher $T_{\rm max}$) are more intensely heated
(higher $E_H$), so that $E_H\propto T_{\rm max}^{\zeta }$ with $\zeta > 0$;
thus we can conjecture that the steep power law dependence in Eq.
\ref{eq29} mostly reflects the dependence of $a_j$ on
$T_{\rm max}$, i.e., that $a_j(T_{\rm max})\propto T_{\rm max}^{-7.25}$ or
even steeper. Equivalently, one may substitute $E_H$ from Eq.
\ref{eq25} into \ref{h_setloop} to obtain
\begin{equation}
h(T_{\rm max})  = \left\{
\begin{array}{ll}
0 & \mbox{for } T_{\rm max} < T_P\\
\\
\displaystyle10^5 \left(\frac{T_{\rm max}}{1.4\times 10^3}\right)^{7/2}
\sum_j
\frac{1}{L_j^2} L_j a_j = 9.7\times 10^{-7} T_{\rm max}^{7/2} \sum_j
\frac{a_j}{L_j} &
\mbox{   for } T_{\rm max}\ge T_P~.
\end{array}\right.
\label{eq30}
\end{equation}
\noindent
This equation shows explicitly the link between the heating budget and the
geometrical factors. It may at first sight appear that there is no
dependence on parameters such as the plasma pressure; these are, however,
involved through the scaling laws and their dependence on $L$.

Appendix B presents an evaluation of the heating of loops with $T_{\rm max}
\approx 2 \times 10^6$K in order to provide the reader with a feeling of
the orders of magnitude involved in this discussion.

\section {Discussion and conclusions}

We have studied the piecewise-integrated emission measure versus
temperature, $EM(T)$, of the solar corona taken as a whole, i.e., as if it
were an unresolved stellar corona.  Using the facts that the corona is
almost entirely composed of steady loops whose height is less than their
pressure scale height and that the emission measure distribution versus
temperature for each static loop is $\propto T^\beta$ (with $\beta
\approx 1.5$), we have found that we can derive the distribution of the
emitting loops versus temperature from the emission measure
distribution $EM(T)$, as well as the heating budget for loops with different
maximum temperature.

Rather than focusing on the energetics of individual loops, this paper 
studies the solar corona taken as a whole, in the same way that stellar
observations treat stellar coronae. In our model of the solar corona,
loops with the same maximum temperature $T_{\rm max}$ are grouped together
as their emission measure distribution as a function of temperature would
be indistinguishable from that of a single loop with the same $T_{\rm max}$
and emission measure equal to the sum of their emission measures. We then
sum all the contributions of such ``equivalence classes" of loops over
$T_{\rm max}$ to obtain the total coronal emission measure distribution.
We show further that the observed total coronal emission measure
distribution versus temperature, $EM(T)$, agrees with the results of such a
model calculation.  Furthermore, we show that if one differentiates the
observed $EM(T)$ with respect to temperature, one obtains the emission
measure for all loops with a given maximum temperature $T_{\rm max}$
present at that time in the corona. This suggests that our method may be
directly applied to the stellar case, in which case loop structures cannot
be resolved individually, and it may represent an alternative to the fitting
with multiple thermal components for those cases in which the emission
measure distribution can be derived.

In a recent work Aschwanden et al. (2000) study coronal loops observed
with TRACE; they claim that these loops appear to be isothermal, i.e. 
very different from the standard scenario given by, for instance, Yohkoh
data, of loops dominated by the balance of heating, radiative losses
and, more important, thermal conduction.

However, the temperature and emission determination by Aschwanden et
al., as the analysis routinely used for TRACE data, is based on narrow
band XUV filter ratio method: each of the narrow bands accepts few
spectral lines which form at rather different temperatures, i.e.
approximately between $10^5$ and $10^7$ K.  As a consequence, the
temperature vs. filter-band-ratio is a multi-valued function, and so
the relevant temperature determination is not-unique.  The solution
chosen, for the data analysis routinely used, has been to force the
$ 195/171$ AA ratio to be (approximately) in the $0.9 - 1.8 \times 10^6$ K
temperature range based on the claim that this is the range of maximum
line visibility at constant emission measure.  As we discuss below,
ignoring the role of emission measure in determining plasma visibility
and the fact that loops can have widely different emission measure
values, can lead to severe errors. 

Indeed, on the basis of coronal loop physics matured over decades of
X-ray observations and modeling, it is known that
hotter loops have a corresponding much
higher pressure and higher emission measure compensating for the
presumed lower visibility, as can be easily seen with hydrostatic loop
models. Since the entire loop density distribution scales with the loop
maximum temperature as $T_{max}^2$, their emission measure measure
should scale as $T_{max}^4$. Taking this feature into account along
with the spectral response of TRACE, loops between $10^6$ K and $10^7$ K
are all almost equally visible, much more visible than presumed in the
TRACE data analysis system.  The plasma detected in all this wide range
is however invariably assumed to be in the $0.9 - 1.8 \times 10^6$ K range: any
thermal structure along the loop is entirely wiped out and all the
loops are claimed to be at the same temperature.

Testa et al. (2001a, 2001b) have analyzed some coronal loops observed with
TRACE, relaxing the above mentioned limitation of temperature and
trying a novel approach to the TRACE data analysis. They have fitted
both the brightness profiles in the $171$ A and $195$ A bands, used for
standard TRACE observations, along the loops {\em and} used the filter
ratio values, after removing the background emission. They find that
some loops appear to be at $5 \times 10^6$ K and not isothermal, while
others are consistent with isothermal but at $10^5$ K, i.e. more alike
some loops observed with SOHO/CDS (Brekke et al., 1997) and S-055 Skylab
(Foukal, 1976).

Furthermore, spectroscopic work by Brosius et al. (1986), Raymond and
Doyle (1981), Jordan (1980), Laming, Drake and Widing (1995) yield a
coronal emission measure typically $\propto T^{3/2}$ and Priest et al.
(2000) obtain a good fit of a coronal loop with a not-isothermal
profile, assuming an energy balance between a heating term, heat
conduction and radiative losses; Priest et al.  claim "strong evidence
against heating concentrated near the loop base", i.e. opposite to one
of the key conclusions of Aschwanden et al. analysis.

Incidentally, Aschwanden et al. (2000) do not show that the loops,
claimed to be isothermal, are statistically important - i.e.  very
frequent - in the overall well proved scenario of hot coronal loops
dominated by heat conduction and definitely not-isothermal.

For all these reasons we have decided not to take into account possible
"isothermal" coronal loops and consider just the well founded coronal
loops scenario based on X-ray photometric observations and UV
spectroscopic studies.

The method presented here is substantially different from that of Maggio \&
Peres (1996), who carry out a ``minimum $\chi^2$" fit of stellar
low-resolution X-ray spectra with one- or two-loop coronal models: here
instead we model the coronal emission measure distribution 
as due to a continuum of loops.

Our results show that, aside from times when flares or other large
transients dominate coronal emission, the temperature dependence of
the emission measure distribution of the entire solar corona can be modeled
as a power law of the form $\propto T^{3/2}$, up to a maximum temperature
which typically lies between $10^6$ K and $2\times 10^6$ K. For $T > T_P$, we
show that the emission measure distribution decreases as a steep power
law in $T$. This decrease of EM(T) for $T > T_P$ can be used to derive the
distributions both of loops and of their heating as a function of $T_{\rm
max}$, the loops' maximum temperature. Finally, we show (on the basis of
Eq. \ref{loop_contrib}) that there are relatively few loops with
maximum temperature well below $T_P$; indeed, we show that the distribution
of emitting loops, derived on the basis of Eq. \ref{loop_contrib}, and
the temperature dependence of the heating rate (Eq.
\ref{heat_budget}) are rather sharp functions of $T_{\rm max}$. An
important (as yet unresolved) issue to what extent this sharp temperature
dependence is real, rather than depending on (either) a somewhat too
idealized model or the unavoidably limited instrumental capabilities of
Yohkoh/SXT (in particular, its spectral resolution). Observations using EUV
spectroscopic data (e.g., Brosius et al.\ 1996) show rather similar
emission measure distributions, suggesting that the steep rise of the loop
distribution is very likely real; nevertheless, it seems extraordinary that
there are very few quasi-steady loops with $T_{\rm max} < 1.7 \times
10^6$ K. In any case, it appears reasonable that the contribution of
quasi-steady loops with maximum temperature below 1 or 2 million degrees to
EM($T$) is indeed very small relative to the hotter corona,
albeit not exactly zero, at the maximum
of the solar cycle.

The largest contribution to the coronal emission measure, to the
coronal radiative losses and to the coronal heating is due to loops
with maximum temperature around the peak of EM$(T)$. The steep decrease
of the solar EM$(T)$ beyond the temperature of maximum EM$(T)$ implies
that loops are fewer and fewer for higher and higher $T_{\rm max}$, and that
less and less total heat is delivered in the hotter loops. This peak
evolves as a function of the solar cycle: From Eq. \ref{loop_contrib},
Eq. \ref{heat_budget}, and from the evolution of EM$(T)$ during the
solar cycle (Paper II; Orlando et al.\ 2000b) we can infer the evolution of
the emitting loop distribution and of the related heating distribution.
Observations show that both distributions become more concentrated at lower
temperatures, and that their maximum value gets smaller during the minimum
of the solar cycle; we also know that the ascending, low temperature,
portion of EM$(T)$ becomes slightly steeper during the solar minimum. The
steepening during the minimum of the cycle may be due to several causes,
including the low Yohkoh/SXT sensitivity to cooler plasma or (more likely)
to the fact that most of the corona is devoid of loops and thus consist
mostly of non-confined plasma. In spite of this, $T_P$ varies only between
$\approx 2 \times 10^6$ K and $ \approx 7 \times 10^5$ K during the solar
cycle.

On the other hand EM($T$), $T_P$ and, therefore, the distribution of
loop emission measure and heating versus $T_{\rm max}$ all appear to be
subject to negligible (if any) changes on time scales of hours or --
sometimes -- days (away from flares). The steadiness of EM$(T)$, and
therefore of the distributions of the loops and of the heating, over a
few hours (Orlando et al.\ 2000b) seems to imply an analogous
steadiness of the heating budget of the entire confined solar corona,
with only slow and gradual changes.  It is not clear why this is so.
One possibility is that this observational characteristic is just a
consequence of summing randomly distributed contributions over the
entire corona, so that it just reflects the effects of statistical
temporal smoothing rather than true steadiness of the entire corona.
Another possibility is that the global externally supplied energy
budget of the corona is actually steady. Either possibility would have
major implications for the heating process of the corona.

Finally, we note that it is straightforward to see from Eq. \ref{Nfqeqn}
that any flattening, or even increase, at a given temperature $T^*$ of the
emission measure EM$(T)$ along its descending portion would provide evidence
for another preeminent set of loops, and related heating, with $T_{\rm
max} \approx T^*$. In particular, a doubly-peaked EM($T$) (as the one found,
for instance, by Griffith \& Jordan 1998) implies a doubly peaked loop
distribution.  In the light of our results, the findings of Griffiths \&
Jordan (1998) are indicative of two distinct distributions of
steady loops peaking at two rather different temperatures, one at several
$10^6$ K and the other at a few $10^7$ K. Alternatively, it may be that
steady coronal emission is accompanied in some (very active) stars by a
relatively large number of simultanteous flares whose light curves
overlap in such a way that they mask much -- if not all -- of the
variability (e.g., Giampapa et al.\ 1996; Paper II); see Reale et al.\
(2000) and Peres (2000).

Our method allows us -- in principle -- to infer the global distribution of
the heating versus the loop maximum temperature; this information can, in
turn, provide constraints on the heating mechanism of the corona and on the
global coronal energy budget. On the practical level, however, it is far
from easy to derive a relationship between the loops' average volumetric
heating rate and the loop temperature as one needs to know the fraction of
the solar surface covered by the footpoints of the loops with given maximum
temperature $T_{\rm max}$.

As for future applications, we plan to apply extensively the
diagnostics presented here to the EM$(T)$ of the whole solar corona at
several points during the solar cycle, using the large data set collected
with Yohkoh/SXT. It is also possible to apply the diagnostics to
selected portions of the corona, e.g., Yohkoh/SXT data on single active
regions, or to SERTS observations (Brosius et al.\ 1996). The diagnostics can
also be modified very easily for application to the diffential emission
measure distribution, rather than to the piecewise-integrated emission
measure treated here. For obvious reasons, our analysis is well suited
to stellar observations since it uses a global coronal characteristics,
EM($T$), readily obtained by a spectroscopic coronal observation. We foresee
an immediate application to EUVE stellar observations (Drake et al.\ 2000);
and we also plan to apply it to data obtained from the spectroscopic X-ray
instruments onboard the Chandra and XMM-Newton X-ray astronomy missions. It
will be particularly instructive to compare the results for stars of different
activity levels with the analogous solar results.

\acknowledgments We acknowledge useful suggestions and comments from S.
Serio and E. Franciosini, and useful discussions with C. Kankelborg.
We also acknowledge useful suggestions from an anonymous referee.
This work has been partially supported by the Agenzia Spaziale Italiana
and by the Italian Ministero dell'Universit\`a e della Ricerca
Scientifica e Tecnologica.

\appendix
\section{Deriving the loop population from EM$(T)$}

Our task is to derive $N~f(T_{\rm max})~em(T_{\rm max})$ from the observed
EM$(T)$. On the basis of Eq. (12),

\begin{eqnarray}
\frac{d~\mbox{EM}(T)}{dT} =~~\beta ~T^{\beta -1}
~~\int_T^\infty ~~ dT_{\rm max}~N~f(T_{\rm max})~em(T_{\rm max})
~(T_{\rm max})^{-\beta}~+
\nonumber
\end{eqnarray}

\begin{eqnarray}
~~~~~~~~~~~+ T^{\beta}
~~\frac{d}{dT} \int_T^\infty ~~ dT_{\rm max}~N~f(T_{\rm max})~em(T_{\rm
max}) ~(T_{\rm max})^{-\beta}
\nonumber
\end{eqnarray}
\begin{eqnarray}
= \beta ~\frac{\mbox{EM}(T)}{T} - N~f(T)~em(T) ~~~~~~~~~~~~~~~~~~~~
\nonumber
\end{eqnarray}

\noindent
since the last term is 

\begin{eqnarray}
\lefteqn{T^{\beta} \lim_{\Delta T \rightarrow 0} \frac{1}{\Delta
T}\left[\int_{T+\Delta T}^{\infty} ~ dT_{\rm max}~N~f(T_{\rm
max})~em(T_{\rm max}) ~(T_{\rm max})^{-\beta} -
\right.}
\nonumber \\
 & \displaystyle - \left. \int_{T}^{\infty} ~ dT_{\rm max}~N~f(T_{\rm
max})~em(T_{\rm max}) ~(T_{\rm max})^{-\beta} \right]  \nonumber \\
 & \displaystyle ~~~~~= T^{\beta} \lim_{\Delta T \rightarrow 0}
\frac{1}{\Delta T} \int_{T}^{T+\Delta T} ~ dT_{\rm max}~N~f(T_{\rm
max})~em(T_{\rm max}) ~(T_{\rm max})^{-\beta}
\nonumber
\\
 & \displaystyle = T^{\beta} \lim_{\Delta T \rightarrow 0} \frac{1}{\Delta
T} N~f(T)~em(T) T^{-\beta} \Delta T ~~~~~~~~~~~~~~~~~~~~~~~
\nonumber
\\
 & \displaystyle = N~f(T)~em(T) ~,~~~~~~~~~~~~~~~~~~~~~~~~~~~~~~~~~~~~~~~~~~~~~~~
\nonumber 
\end{eqnarray}

\noindent
and therefore

\begin{equation}
N~f(T)~em(T)= \beta ~\frac{\mbox{EM}(T)}{T}~~-
~~\frac{d~\mbox{EM}(T)}{dT}~.
\label{eq_app1} 
\end{equation}
This expression connects the distribution function of loop maximum
temperatures to the global emission measure distribution. In the
specific case of a $\beta = 3/2$ power law, we obtain instead of Eq.\
\ref{eq_app1}

\begin{equation}
N~f(T)~em(T)= ~~\frac{3}{2}~\frac{\mbox{EM}(T)}{T}~~-
~\frac{d~\mbox{EM}(T)}{dT}~.
\end{equation}

%\appendix
\section{Example: the heating budget of loops with $T_{\rm max} \approx 2
\times 10^6$K}

In this appendix, we evaluate the heating budget of loops with $T_{\rm
max}$ in the range of $2 \times 10^6$K, using a few simple assumptions; our
aim is to provide the reader with a feeling of the orders of magnitude
involved.

As a first simplification, we note that since loops with $T_{\rm max}
\approx 2 \times 10^6$K largely dominate the emission measure distribution,
their total emission measure $\epsilon$ is comparable to the maximum value
of $EM(T)$ and therefore we shall assume $\epsilon \approx 10^{49}$
cm$^{-3}$. The
emission measure in an equivalence class of loops with maximum
temperature $T_{\rm max}$ (all with the same length $L$ and, therefore,
pressure $p$) can then be approximated as
$\epsilon= ~ n_p^2 V= ~  p^2 V /( 2 k_B T_{\rm max})^2$; $n_p$ is the proton number
density, V is the loops' total volume, and $k_B$ is the Boltzmann constant;
according to the RTV loop scaling laws, $T_{\rm max}^3 = (1.4 \times 10^3)^3 ~p
~L$, and therefore $n_p^2= T_{\rm max}^4 / (b L^2)$, where $b \approx 5.7 \times
10^{-13}$. Thus, upon substitution, we find that the emission measure is
given by the expression $\epsilon= T_{\rm max}^4 V / (b L^2)$ and $V= \epsilon~ b~
L^2/ T_{\rm max}^4$.

However, loops with the same temperature may have different, even very
different, lengths. Let us consider, then, for simplicity just two values
of $L$, $10^9$ cm and $10^{10}$ cm, the first representative of active
regions loops, the other of more quiet and extended loops but still shorter
than the pressure scale height; their total emission measure is
$\epsilon_1 + \epsilon_2= ~\epsilon$ and their total volume is $V= V_1 +
V_2= b~ ( EM_1 L_1^2 + EM_2 L_2^2)/T_{\rm max}^4 $ (where, from here on, all
quantities with suffixes 1 and 2 pertain to, respectively, the first and
the second set of loops). Then if we assume that $V_1 = ~ V_2$, and since
$L_1= 0.1 L_2$, we are led to the result $\epsilon_1= ~100 ~ \epsilon_2$. 
If we insert numerical values for the various known quantities, we then
obtain $V \approx 10^{30}$ cm$^3$.

Furthermore we may assume that loops have an aspect ratio $L/r \approx 10$
(a typical value of most loops observed) where r is the cross-sectional
radius of the loop. The volume of each loop is then $v= ~2 \pi r^2 L=~
2 \times \pi 0.1^2 L^2 \times L = ~~ 0.02 \pi L^3$. On the basis of these
considerations we would then have $n_1$ loops for the first set and
$n_2$ for the second, such that $n_1= ~V/~2 v_1 \approx 8 \times 10^3$
loops and $n_2 \approx 8$ loops.

Under the same assumptions, the sum in Eq. (38) can be re-written in
the form
\begin{equation}
\Sigma a_j/L_j=
\Sigma \pi 0.01 L_j ~,
\end{equation}
so that the contribution of all loops with the same length
within the $T_{\rm max}$ equivalence class would amount to $n \pi
\times 0.01 L$. Thus, for the first of loops considered above, $n_1 \pi
\times 0.01 L_1= 8 \times 10^3 \pi \times 0.01 \times 10^9 = ~ 2.5 \times
10^{11}$ and $n_2 \pi \times 0.01 L_2= 8 \pi \times 0.01 \times
10^{10}= 2.5 \times 10^9$. Evidently under these assumptions the set 1 of
loops, i.e., those with $L= 10^9$cm, have a significantly larger sum, and
therefore dominate the heating budget of the $T_{\rm max}$ equivalence
class. Thus, from Eq. (38) we have $h(2 \times 10^6 K)\approx 10^{-6}
\times 11.3 \times 10^{21} \times ~ 2.5 \times 10^{11} erg \approx 28.3 \times
10^{26}$ erg.

\clearpage
\large
\bigskip
\begin{center}
{\bf References}
\end{center}
%%%\end{document}
\small
\begin{description}

\item Aschwanden, M.J., Nightingale, R.W., \& Alexander, D. \ \ 2000, \apj,
541, 1059

\item Athay, R.G., 1966, ApJ 145, 784

\item Brekke, P., Kjeldseth-Moe, O., \& Harrison, R. A.\ \ 1997, Solar
Physics, 175, 511

\item Brosius, J.W., Davila, J.M., Thomas, R.J., \& Monsignori--Fossi,
B.C.,\ \ 1996, \apjs, 106, 143

\item Drake, J.J., Peres, G., Orlando, S., Laming, J.M., \& Maggio,
A.\ \ 2000, \apj, 545, 1074

\item Foukal, P.V. \ \ 1976, \apj, 210, 575

\item Giampapa, M. S., Rosner, R., Kashyap, V., Fleming, T. A., Schmitt,
J. H. M. M., \& Bookbinder, J. A.\ \ 1996, \apj, 463, 707

\item Griffiths, N.W., \& Jordan, C., 1998, ApJ 497, 883

\item Jordan, C., 1980, A\&A 86, 355

\item Kaastra, J.S.\ \ 1992, An X-Ray Spectral Code for Optically Thin
Plasmas (Internal SRON-Leiden Report, updated version 2.0)

\item Laming, J.M., Drake, J.J., \& Widing, K.W. \ \ 1995, \apj 443, 416

\item Maggio, A., \& Peres, G., 1996, A\&A 306, 563

\item Malinovsky, M., \& Heroux, L., \ \ 1973, \apj 181, 1009

\item Mewe, R., Kaastra, J.S., \& Liedahl, D.A.\ \ 1995, Legacy, 6, 16

\item Mewe, R., Lemen, J.R., \& van den Oord, G.H.J.\ \ 1986, A\&AS, 65,
511

\item Orlando, S., Peres, G., \& Reale, F.\ \ 2000a, \apj 528, 524 [Paper
I]

\item Orlando, S., Peres, G., \& Reale, F.\ \ 2000b, in preparation

\item Peres, G., 2000, Proc. of the ``X--Ray Astronomy 2000'' meeting,
PASP, in preparation

\item Peres, G., Orlando, S., Reale, F., Rosner, R., \& Hudson, H.\ \
1997, in "Observational Plasma Astrophysics: Five Years of Yohkoh and
beyond", Watanabe, T., Kosugi, T., Sterling, A.C. eds., Kluwer, p. 29

\item Peres, G., Orlando, S., Reale, F., Rosner, R., \& Hudson,
H.\ \ 2000, \apj, 528, 537-551 [Paper II].

\item Priest, E.R., Foley, C.R., Heyvaerts, J., Arber, T.D., Mackay, D.,
Culhane, J.L., \& Acton, L.W. \ \ 2000, \apj, 539, 1002

\item Raymond, J.C., \& Doyle, J.G. \ \ 1981, \apj, 247, 686

\item Reale, F., Orlando, S., \& Peres, G., et al.\ \ 2000, in
preparation

\item Rosner, R., Tucker, W.H., \& Vaiana, G.S., 1978, ApJ 220, 643

\item Sheeley, N., \& Golub, L.\ \ 1979, Sol. Phys. 63, 119.

\item Shimizu, T., \& Tsuneta, S.\ \ 1997, ApJ 486, 1045.

\item Testa, P., Peres, G., Orlando, S., \& Reale, F., \ \ 2001 a, in
Solar Encounter: the First Solar Orbiter Workshop, ESA Special
Publication SP-493, in press

\item Testa, P., Peres, G., Orlando, S., \& Reale, F., \ \ 2001 b, in
preparation

\item Title, A., \& Schrijver, K.\ \ 1998, ASP Conf. Ser. 154, The
Tenth Cambridge Workshop on Cool Stars, Stellar Systems and the Sun,
Edited by R. A. Donahue and J. A. Bookbinder, p.345

\item Vaiana, G.S., Krieger, A.S., \& Timothy, A.F.\ \ 1973, Sol.\ Phys,
32, 81.

\item Vaiana, G.S., \& Rosner, R.\ \ 1978, ARA\&A,16, 393-428.

\item Ventura, R., Maggio, A., \& Peres, G., 1998, A\&A 334, 188

\item Vesecky, J.F., Antiochos, S.K., \& Underwood, J.H.\ \ 1979, ApJ,
233, 987

\end{description}
\clearpage

\figcaption{A simple example of summing the piecewise integrated
emission measure distribution of various loop equivalence classes (six
in this case), of different temperature maxima.  The thin lines yield
the $T^{3/2}$ contribution of each class of loops having the same
$T_{\rm max}$ and the dotted vertical lines mark the related $T_{\rm
max}$ values.  The solid histogram is the sum of all the equivalence
classes and is $\propto T^{3/2}$ up to the lowest $T_{\rm max}$.  The
inset shows the assumed $N f(T_{\rm max}) em(T_{\rm max})$.
\label{EMsum}}

\figcaption{Piece-wise integrated emission measure distribution for the
whole corona, observed with Yohkoh/SXT on 6 Jan 92, at 21:45 UT (top),
close to the maximum of the solar cycle.
\label{EM1}}

\figcaption{$N f(T_{\rm max}) em(T_{\rm max})$ derived from the EM$(T)$
of Fig.  \ref{EM1}. Diamonds, error bars and solid connecting line are
the results of the Monte-Carlo calculation
discussed in the text: diamonds mark the average values, small error tics
on the error bars
enclose the 68\% of the values, the large ones the 90\%.
The dashed line shows the result in the alternative calculation, approximating
the ascending part up to the maximum with a $T^{3/2}$ power law
and the descending part with a $T^{-3.5}$ power law, and then
applying Eq. \ref{loop_contrib} to the two power laws.  \label{Nfq}}

\figcaption{The function $h(T_{\rm max})$, i.e. the heating budget of
each equivalence class corresponding to maximum loop temperature $T_{\rm
max}$.
\label{h(T)}}

\figcaption{The function $H(t)$, i.e., the heating budget of
the whole corona, integrated up to
temperature t. The figure shows that most of the contribution occurs
for $t\leq T_P$; $T_P$ is where the slope visibly changes.
\label{H(t)}}

\figcaption{$\sum_j E_{Hj}^{1/2}~a_j$ for loops with maximum
temperature $T_{\rm max}$, derived from the EM$(T)$ of Fig. \ref{EM1}.
\label{Ea}}

\newpage
\begin{figure}
\figurenum{1}
\plotone{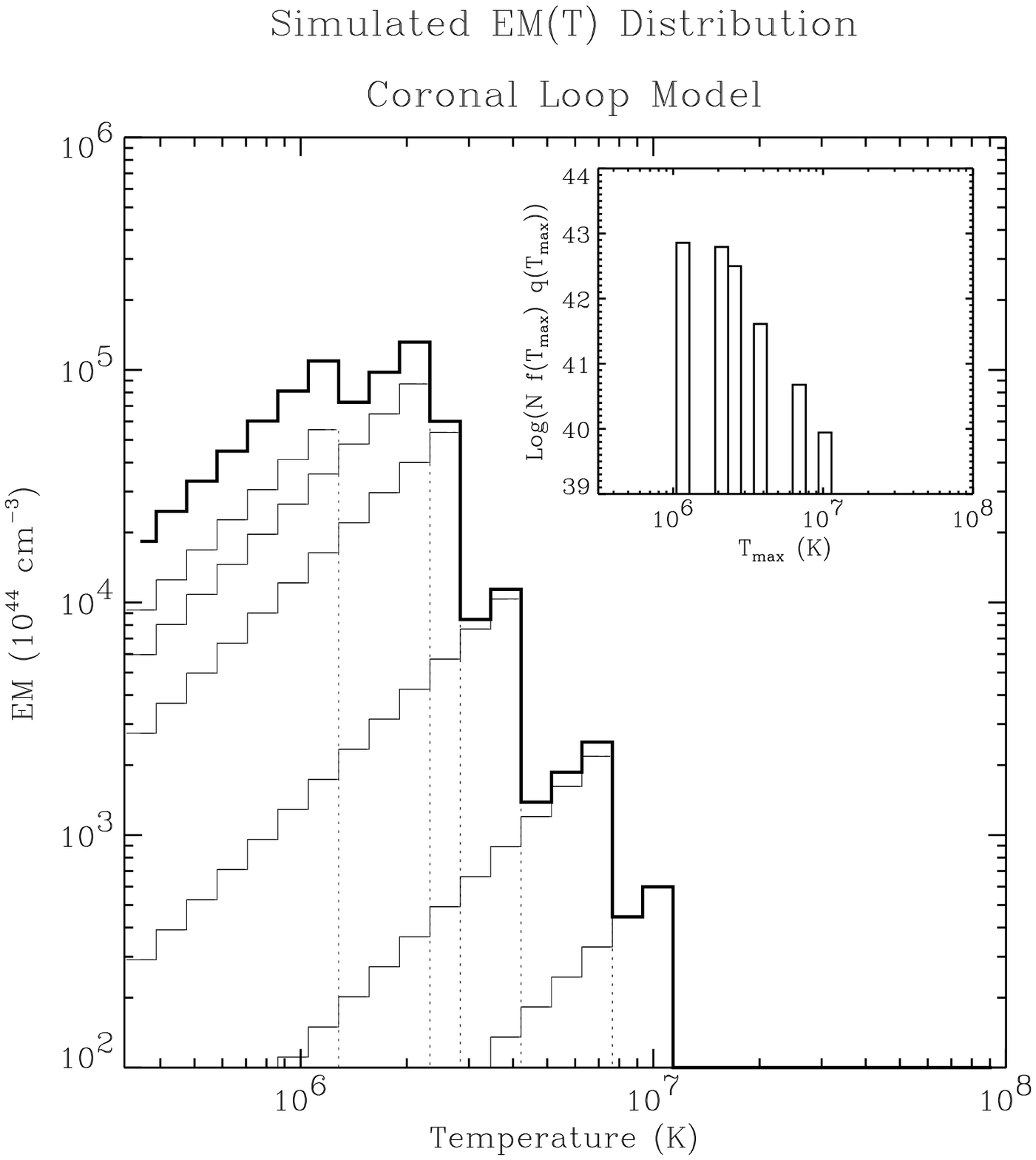}
\caption{}
\end{figure}

\begin{figure}
\figurenum{2}
\plotone{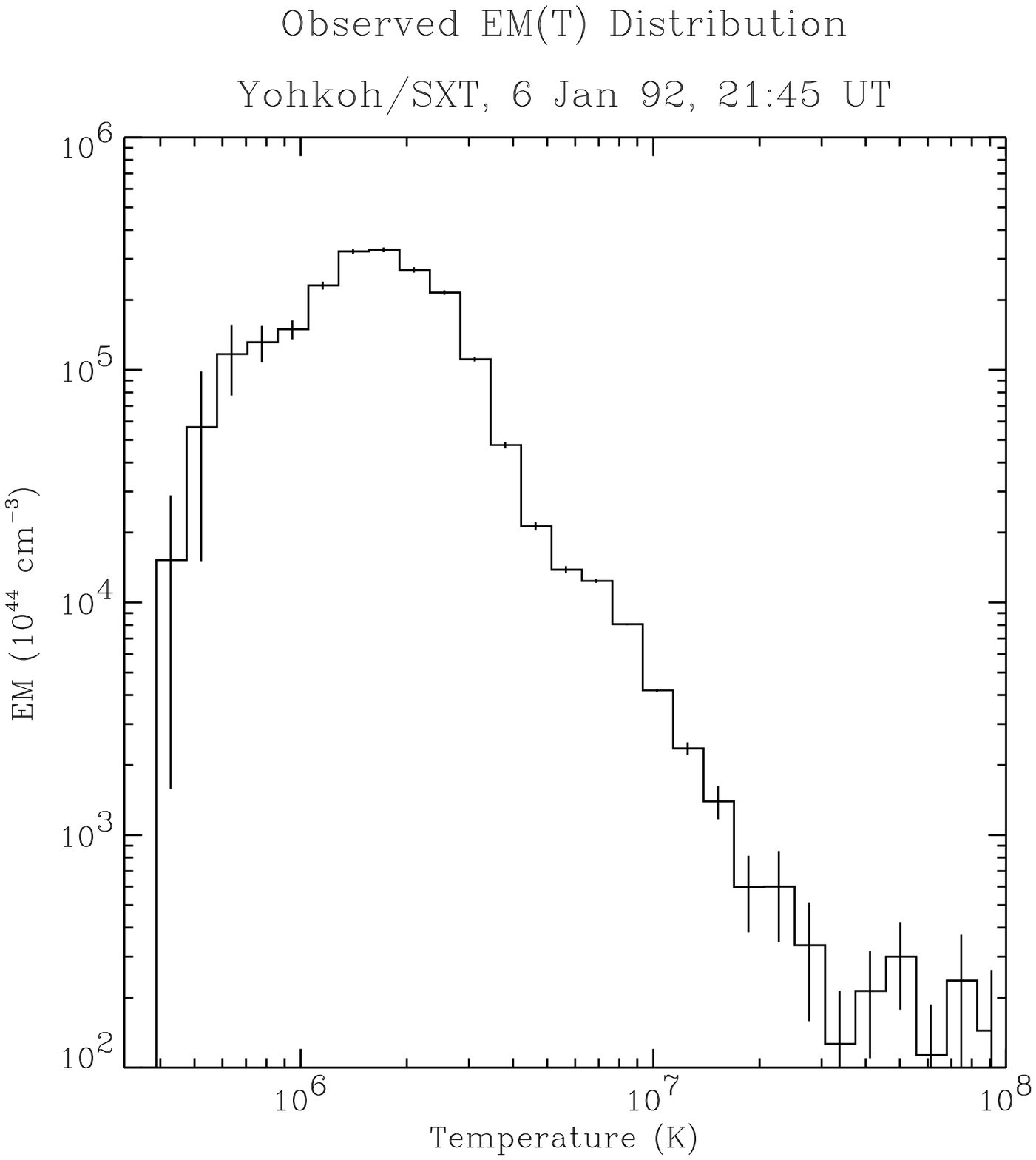}
\caption{}
\end{figure}

\begin{figure}
\figurenum{3}
\plotone{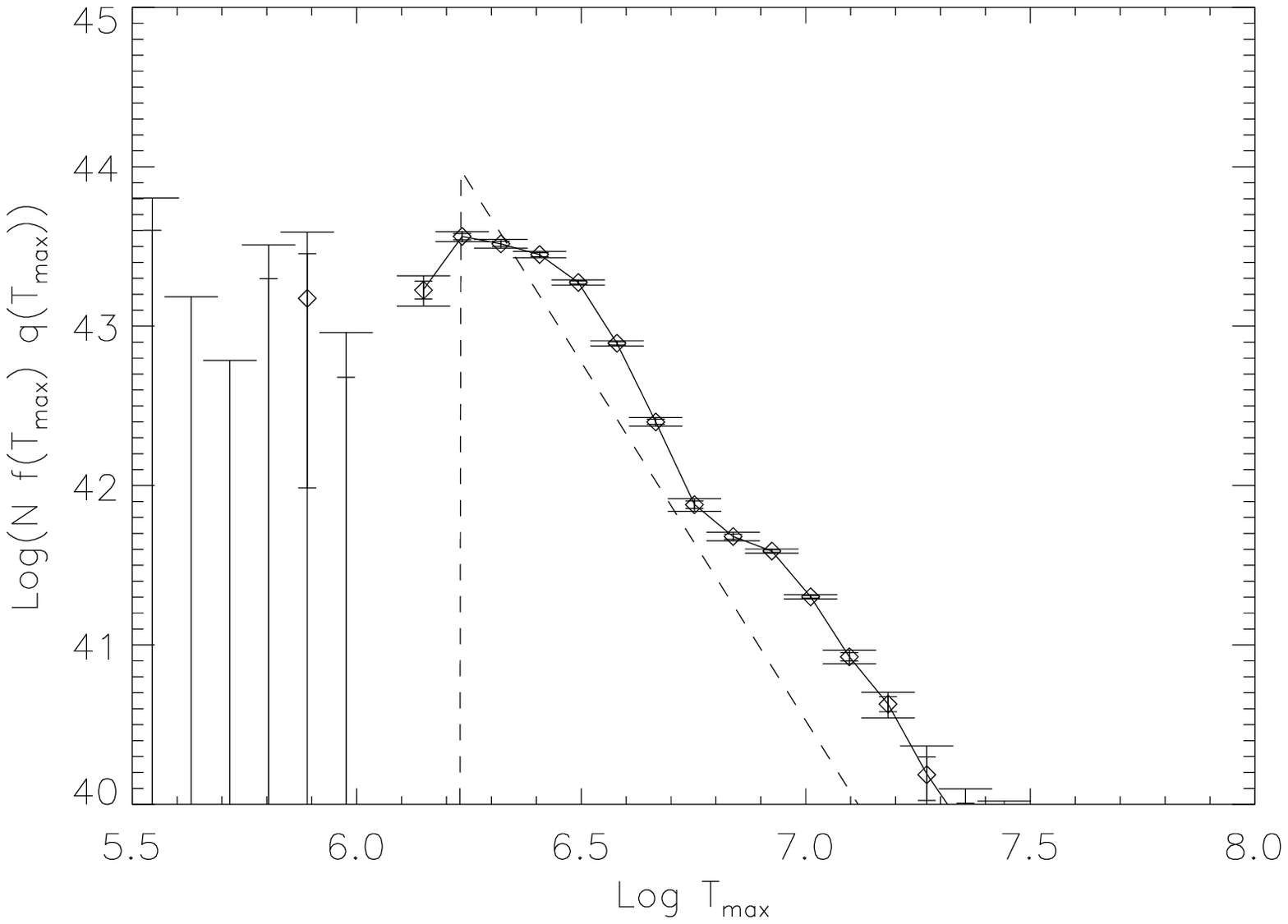}
\caption{}
\end{figure}

\begin{figure}
\figurenum{4}
\plotone{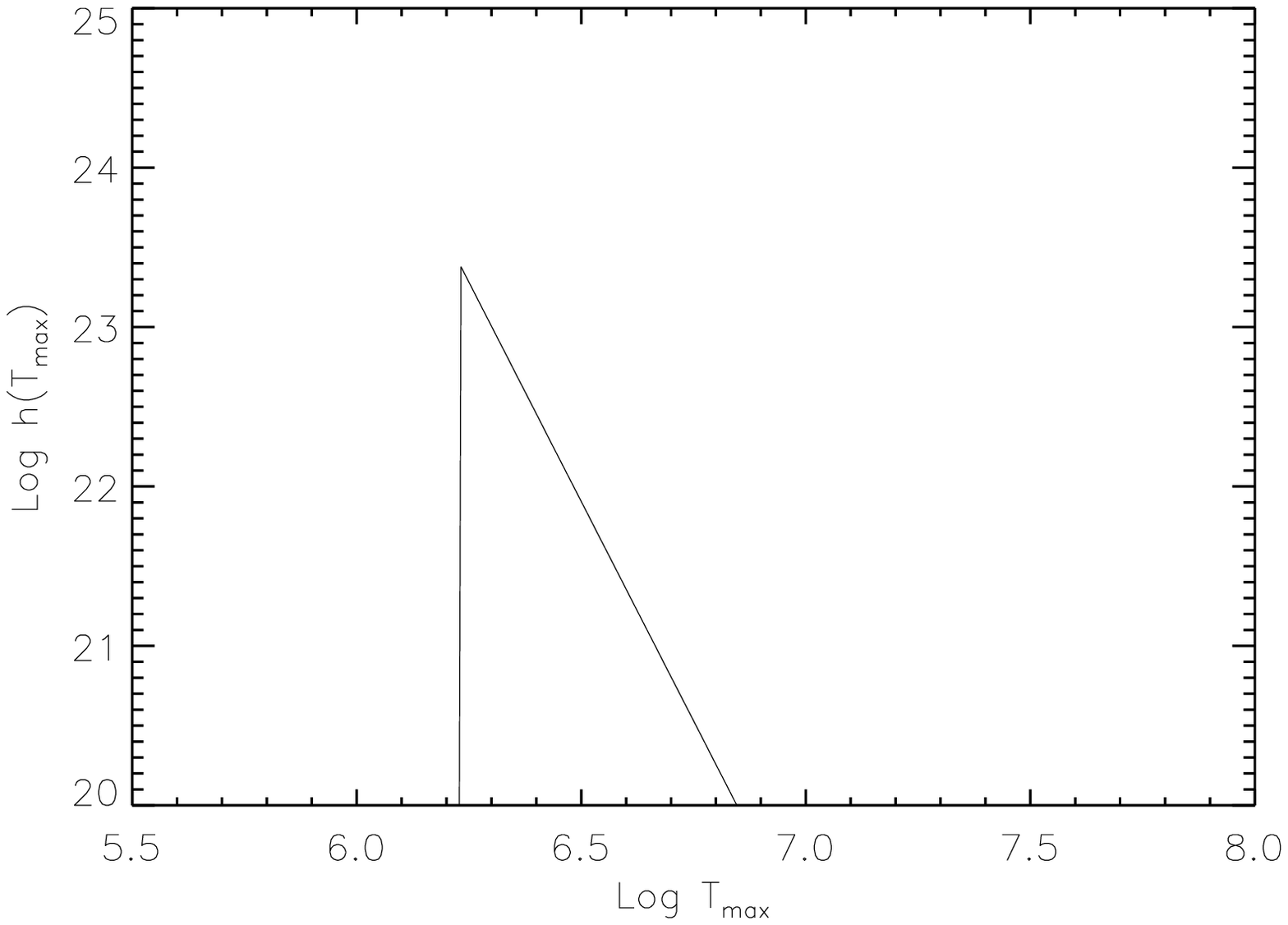}
\caption{}
\end{figure}

\begin{figure}
\figurenum{5}
\plotone{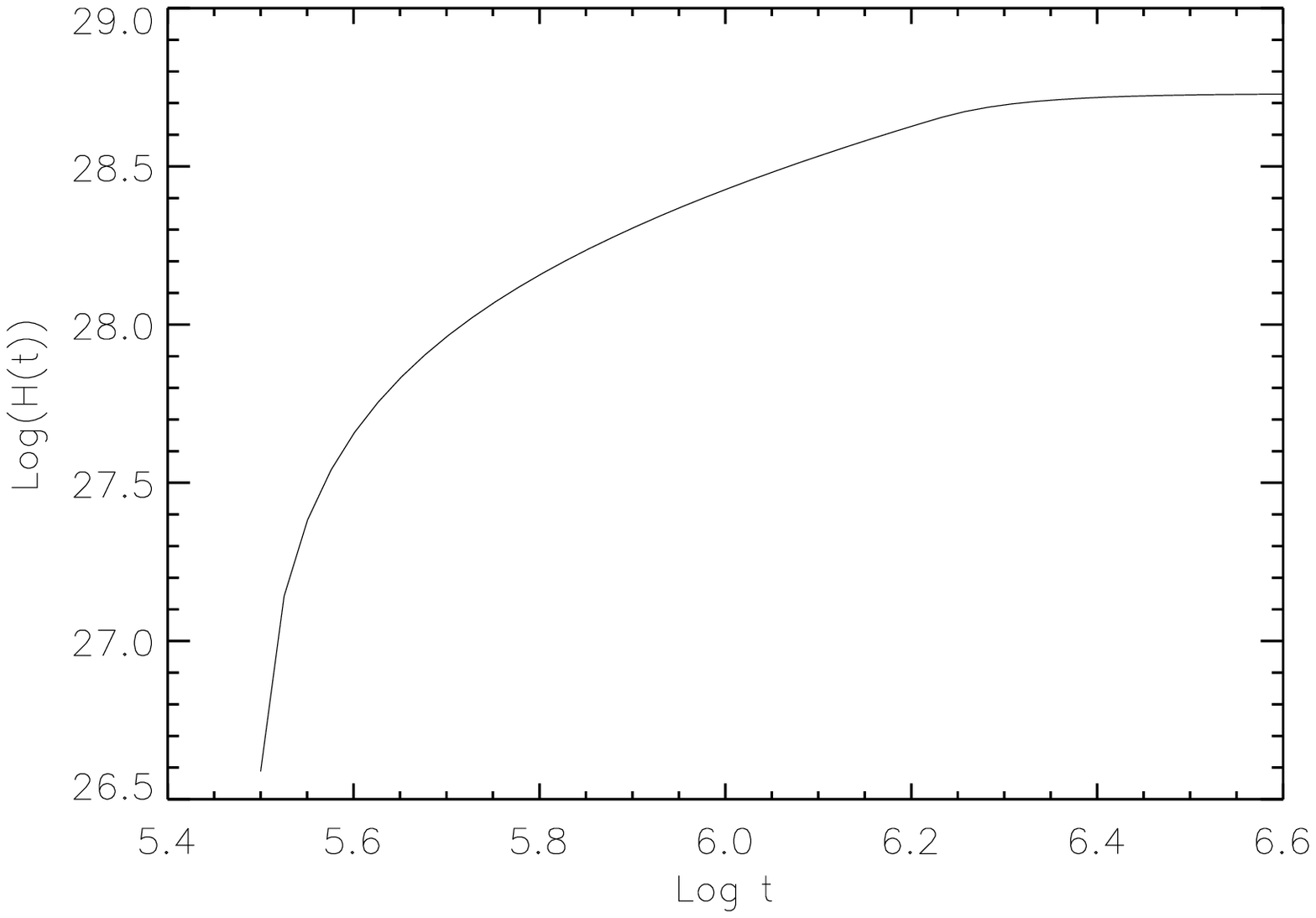}
\caption{}
\end{figure}

\begin{figure}
\figurenum{6}
\plotone{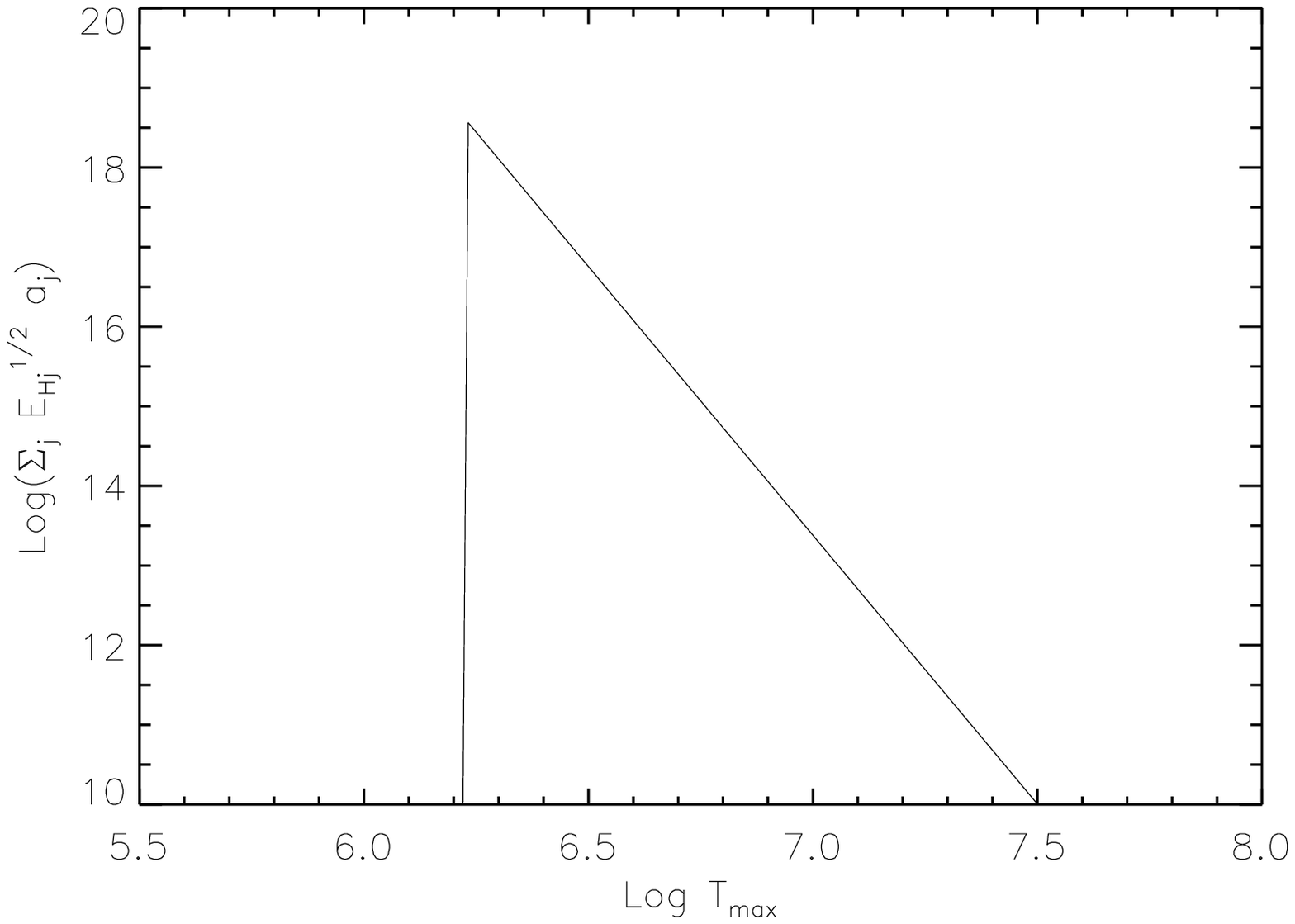}
\caption{}
\end{figure}

\end{document}